\documentclass[a4paper,11pt]{article}
\usepackage{pos}
\usepackage{xspace}
\usepackage{hyperref}
\usepackage{fontawesome}
\usepackage{amsmath}

\newcommand{\capgen}{\texttt{Capt'n General}\xspace}
\newcommand{\DarkMesa}{\texttt{DarkMesa}\xspace}
\newcommand{\mesa}{\texttt{MESA}\xspace}
\newcommand{\githubmaster}{\href{https://github.com/aaronvincent/captngen}{\faGithub}\xspace}

\title{Capt'n General: A generalized stellar dark matter capture and heat transport code}

\author[a,b]{Neal Avis Kozar}
\author[c]{Ashlee Caddell}
\author[d,a,b]{Luke Fraser-Leach} 
\author[c]{\\Pat Scott}
\author[*,a,b,e]{Aaron C. Vincent}
\emailAdd{*aaron.vincent@queensu.ca}

\affiliation[a]{Department of Physics, Enigneering Physics and Astronomy, Queen's University, Kingston, ON, K7L 3N6, Canada}
\affiliation[b]{Arthur B. McDonald Canadian Astroparticle Physics Research Institute, Department of Physics, Engineering Physics and Astronomy, Queen's University, Kingston ON K7L 3N6, Canada}
\affiliation[c]{School of Mathematics and Physics, The University of Queensland, St. Lucia, Brisbane, QLD 4072, Australia}
\affiliation[d]{Department of Physics and Atmospheric Science, Dalhousie University, Coburg Road Halifax, B3H1A6, Canada}

\affiliation[e]{Perimeter Institute for Theoretical Physics, Waterloo ON N2L 2Y5, Canada}

\abstract{\capgen is a  FORTRAN90 standalone package that can be used to compute the capture and heat transport of dark matter in stars. It can compute capture rates for constant, velocity and momentum-dependent DM-nucleon elastic scattering cross sections, as well as non-relativistic effective operator interactions. \capgen can be interfaced with the GAMBIT global fitting codebase as well as stellar evolution simulation codes such as \mesa. \githubmaster }

\begin{document}
\maketitle

\section{Introduction}
It has been known for several decades that if dark matter consists of weakly interacting massive particles (WIMPs) in the $\gtrsim 3$ GeV mass range, the presence of elastic DM-nucleus interactions leads to capture of sizeable populations of DM in the Sun and other stellar bodies. If the DM can self-annihilate to Standard Model (SM) particles that decay to neutrinos, these can be detected at Earth and provide strong constraints -- or a smoking gun signal -- of particle DM. Conversely, if the DM is Asymmetric (ADM), annihilation is suppressed and the DM may act as an efficient \textit{heat conductor} \cite{Steigman78}, leading to observable effects in the Solar fusion neutrino fluxes and helioseismology \cite{Bottino02}. In the case of stars exposed to a higher DM flux (e.g. closer to the galactic centre), lifetimes and evolutionary tracks can also be altered \cite{SalatiSilk89,Scott09,Zentner:2011wx,Taoso08,Raen:2020qvn}.

Here, we present \capgen, a FORTRAN90 code which computes the DM capture rate via the integral over solar radius $r$ and DM halo velocity $u$:
\begin{equation}
C_\odot(t) = 4\pi \int_0^{R_\odot} r^2 \int_0^\infty \frac{f_\odot(u)}{u} \, w \Omega(w,r) \, d u \, d r \, , 
\end{equation}
where $w(r) = \sqrt{u^2 + v^2_{esc,\odot}(r)}$ is the DM velocity at position $r$, and
\begin{align}
 \Omega(w) &= w \sum_i n_i(r,t) \frac{\mu_{i,+}^2}{\mu_i}\Theta\left(\frac{\mu_i v^2}{\mu_{i,-}^2} - u^2 \right) \nonumber \int_{m_\chi u^2/2}^{m_\chi w^2 \mu_i/2\mu_{i,+}^2} \frac{d\sigma_{i}}{dE_R} \, d E_R \,
\end{align}
is the probability of scattering from velocity $w$ to a velocity less than the local Solar escape velocity $v_{\mathrm{esc},\odot}(r)$.  ${d\sigma_{i}}/{dE_R}$ is the DM-\textit{nucleus} scattering cross section, $n_i(r)$ is the number density of species $i$ with atomic mass $m_{N,i}$, and $\mu_i = m_\chi/m_{N,i}$.  \capgen can handle constant spin-dependent (SD) and spin-independent (SI) DM-nucleon cross sections, as well as velocity and momentum-dependent cross sections (as described in \cite{Vincent:2014jia,Vincent:2015gqa,Vincent:2016dcp}) and non-relativistic effective operator (NREFT) interactions \cite{Fitzpatrick13,Catena:2015uha}. 
In addition to capture, \capgen can also compute heat transport rates using the method developed in \cite{GouldRaffelt90a} and expanded \cite{VincentScott2013}. 

We first describe the implementation and use of the \texttt{capgen} code in Sec. \ref{sec:code}. Sec. \ref{sec:capabilities} details the physics capabilities of \capgen. Finally, Sec. \ref{sec:future} details plans for future improvements. 
\section{The \capgen code}
\label{sec:code}
The base function of the \capgen code is to compute the capture rate of halo dark matter in a stellar object, in the optically thin (single-scatter) limit, valid for cross sections well that extend well beyond WIMP regime ($\sigma \sim 10^{-35}$ cm$^2$). The code is written in FORTRAN90, and compiles to produce a shared library object containing necessary functions and subroutines. This is compiled by appropriately modifying the Makefile, and invoking \texttt{make}. Units in the code are cgs, except for input particle masses, which are expressed in units of GeV, and nonrelativistic Wilson coefficients (if used), which have units of GeV$^{-2}$. All capture rates are returned in s$^{-1}$. The code takes some inspiration from the solar capture rate code in DarkSusy \cite{darksusy}, and includes some of the same NIST integration routines.
\subsection{Standalone}
A basic example of how to call \capgen routines is included in \texttt{main.f90}. After compilation of the .so library, the executable can be compiled via \texttt{make gentest.x}. The subroutine \texttt{captn\_init} must be called first, to initialize the Solar (or other stellar) model, and to set the DM halo density and velocity parameters. Capture rates can then be computed for arbitrary DM masses and cross sections. 
\subsection{GAMBIT}
Version 1.0 of \capgen is available as a backend for the GAMBIT \cite{Athron:2017ard} Global and Modular Beyond-the-standard-model Inference Tool (and used in \cite{Athron:2018hpc}), and version 2.1, including NREO Wilson coefficients, will be available by default as of the next GAMBIT release. \capgen can be downloaded and compiled in GAMBIT with the \texttt{make capgen} command, when building other backends. \capgen interfaces with \texttt{DarkBit} \cite{Workgroup:2017lvb}, and can take velocity and momentum-dependent cross sections as inputs, in order to compute the total DM annihilation rate in the Sun, which can then be used for neutrino flux calculations and constraints. 
\subsection{Stellar evolution codes}
In addition to pre-computed Solar model files, \capgen can take as inputs stellar models passed from the \mesa stellar evolution code. DM capture is computed in the same way, and the heat transport routine \texttt{transgen} can compute the transported energy per unit stellar density due to asymmetric DM, and return the outputs to \mesa. \DarkMesa, the interface between \capgen and \mesa will be publicly released in conjunction with an upcoming publication. We also point the reader to Ref. \cite{Raen:2020qvn}, who have provided a similar interface with \mesa, though we note that they make use of a simplified capture algorithm, and only use the Spergel \& Press \cite{Spergel85} transport formalism.

\section{Capabilities}
\label{sec:capabilities}
\subsection{Astrophysical parameters}
\capgen can read in any Solar or stellar model file prepared with the conventions used by the Barcelona group\footnote{\url{https://www.ice.csic.es/personal/aldos/Solar_Models.html}}, noting that the same column order must be followed. \capgen comes with the latest B16 solar models \cite{Vinyoles:2016djt}. It assumes the DM follows a Maxwellian velocity distribution. in the initialization subroutine \texttt{captn\_init}, the user specifies the local DM density in GeV cm$^{-3}$, the stellar velocity, the DM velocity dispersion and the local halo escape velocity in km s$^{-1}$. The capture routines themselves do not account for the geometric limit, for which the DM capture rate is limited by the effective size of the star and the local DM density. This is calculated by invoking the function \texttt{maxcap}.

\subsection{Constant cross sections}
Capture rates for constant DM-nucleon cross sections can be computed with the \texttt{captn\_specific} subroutine, which takes as input the DM mass $m_\chi$, and the respective spin-dependent and spin-independent DM-nucleon cross sections $\sigma_{SD}$, and $\sigma_{SI}$. The SD case simply assumes capture on hydrogen alone; if the spin-dependent effects of heavier elements are required, we recommend using the NREO setup with the corresponding coefficient, $c_4$; see Sec \ref{sec:nreo}, below.

\subsection{Velocity and momentum-dependent capture}
\label{sec:vnqn}
Capture using velocity or momentum-dependent cross-sections, $\sigma = \sigma_0\left(v/v_0\right)^{2n_v}$, $\sigma = \sigma_0\left(q/q_0\right)^{2n_q}$ is handled in terms of a generalized form factor integral (GFFI) as detailed in \cite{Vincent:2015gqa}. Computing capture rates based on such interactions is the purpose of the subroutine \texttt{captn\_general}. The user must specify a DM mass, the power $n_v$ or $n_q$, a reference velocity $v_0$ or momentum $q_0$, and the reference cross section $\sigma_0$. This routine uses a Helm form factor, and only allows nonzero $n_q$ \textit{or} $n_v$.
\subsection{Non-relativistic effective operators}
\label{sec:nreo}
A major development in version 2.0 and later of \capgen is the inclusion of the non-relativistic effective operator (NREO) parametrization. After initialization with \texttt{captn\_init\_oper}, \texttt{captn\_oper} works much in the same way as the \texttt{captn\_general} subroutine, except that it takes as input an array of Wilson coefficients corresponding to the effective operators $\mathcal{O}_k$, $k = 1-15$, first tabulated by \cite{Fitzpatrick13}. These consist of possible scalar combinations of the $\vec S_N$, $\vec S_\chi$, $\vec q$ and $\vec v_\perp$ operators. These are respectively the nucleon spin, the DM spin, the exchanged momentum, and the relative velocity component orthogonal to $\vec q$. The couplings to each operator are $c_k^\tau$, where $\tau = 0$ corresponds to a (weak) isoscalar coupling, while $\tau = 1$ is the isovector coupling.\footnote{The couplings to protons and neutrons are respectively $c_k^p = \left(c_k^0 + c_k^1\right)/2$ and $c_k^n = \left(c_k^0 - c_k^1\right)/2$. } In the NREO formalism, the differential DM-nucleus cross section is proportional to 
\begin{equation}
    \sum_{\tau, \tau^{'}, B} R_B^{\tau\tau^{'}}\left(v_T^{\perp 2}, \frac{q^2}{m_N^2}\right) W_B^{\tau\tau^{'}}(E_R).
\end{equation}
The DM response functions $R_B^{\tau\tau^{'}}$ are functions of the $c_k^\tau$ corresponding to different nuclear response operators $B$. \capgen uses the nuclear response functions $W_B^{\tau\tau^{'}}(E_R)$ for the 16 most abundant elements in the Sun, calculated by \cite{Catena:2015uha}. These are parametrized as polynomials in $y$ times $e^{-2y}$, where $y = (bq/2)^2$, and, $b = \sqrt{41.467/(45A^{-1/3} - 25A^{-2/3})}$ fm \cite{Catena:2015uha}. As this looks like a polynomial times a Helm form factor, the integral over $E_R = q^2/2m_\chi$ can be performed using the same GFFI method as in the $q^n$ case \cite{Vincent:2015gqa}. 

\begin{figure}
    \centering
    \includegraphics[width=0.5\textwidth]{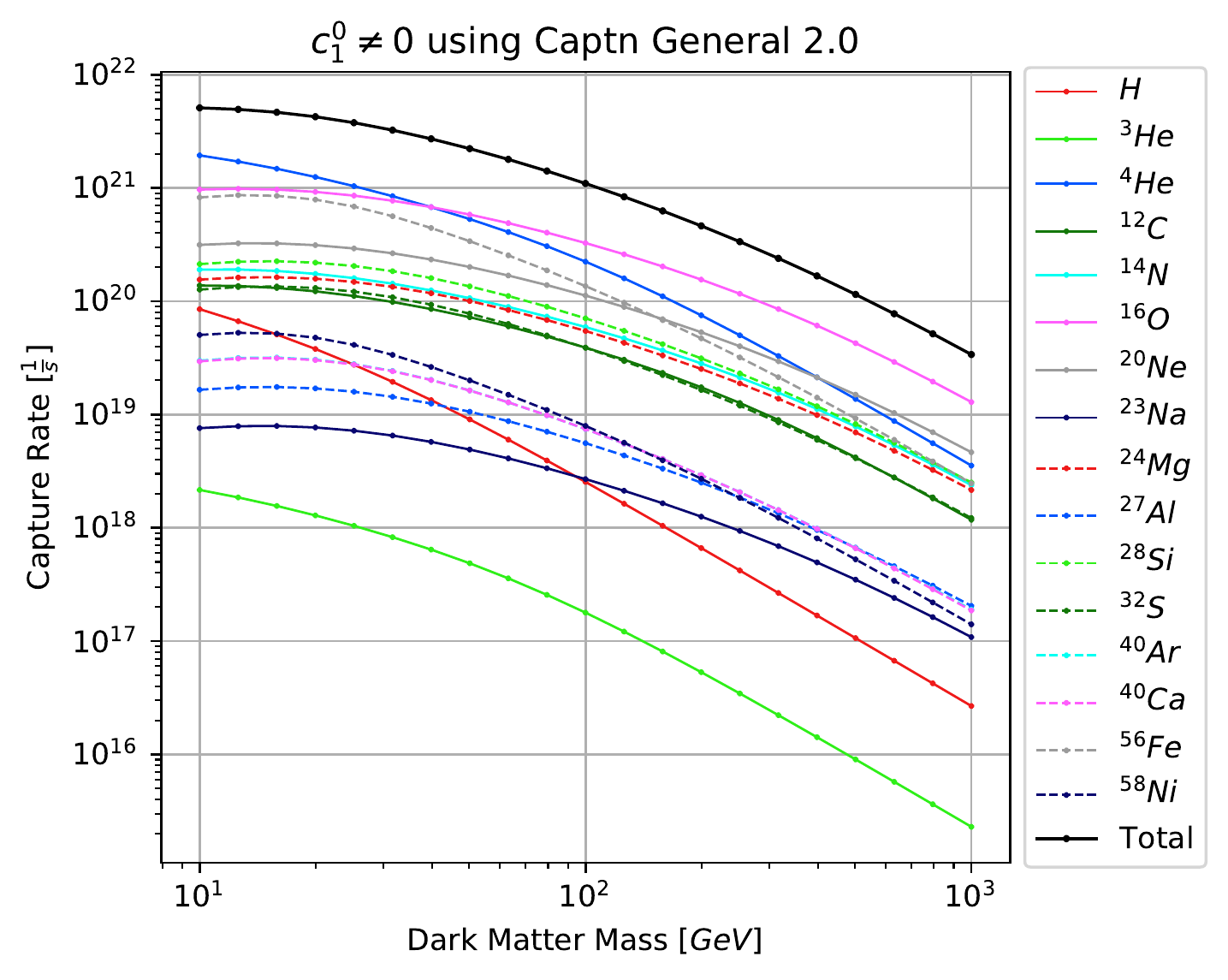}\includegraphics[width=0.5\textwidth]{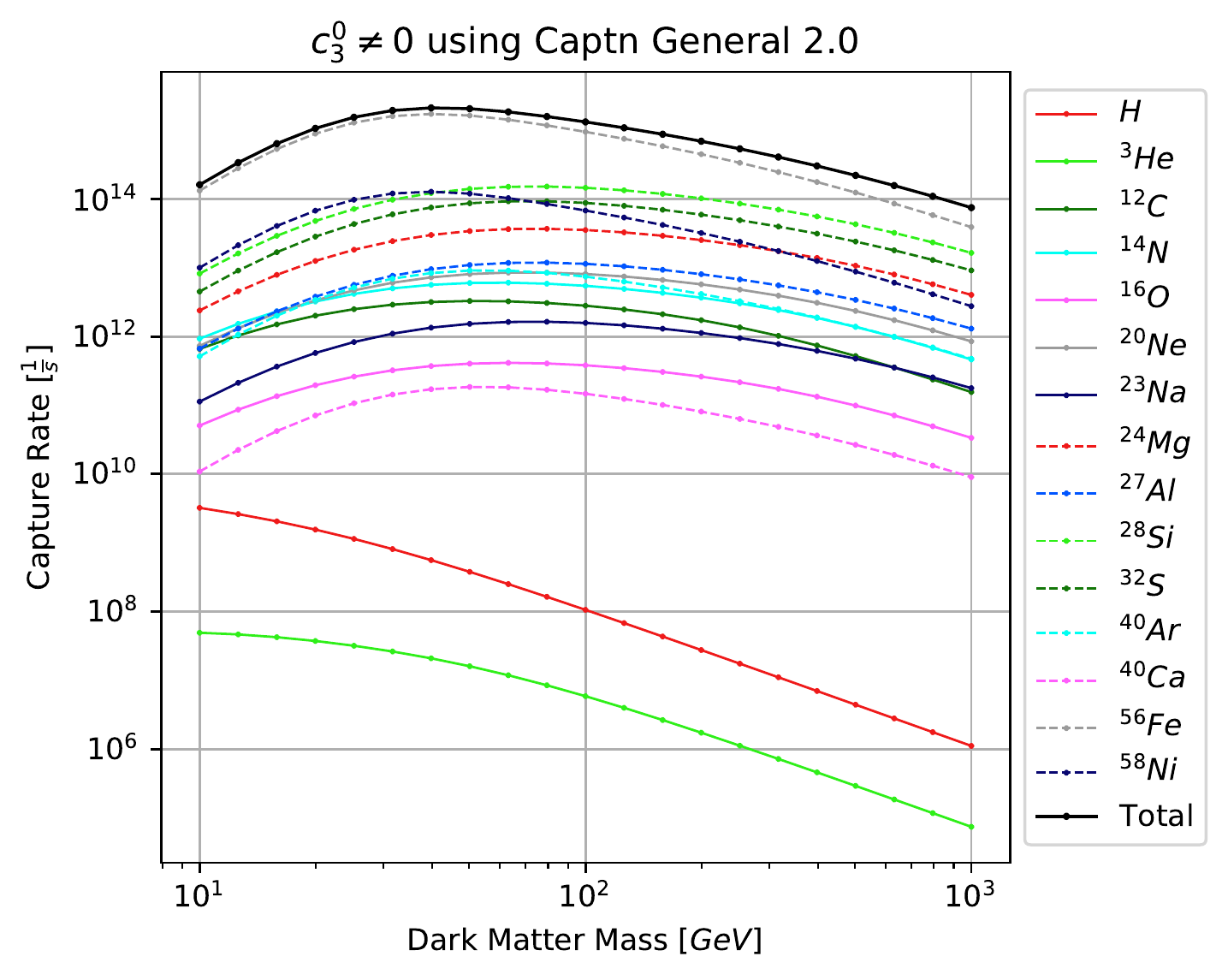}\\
    \includegraphics[width=0.5\textwidth]{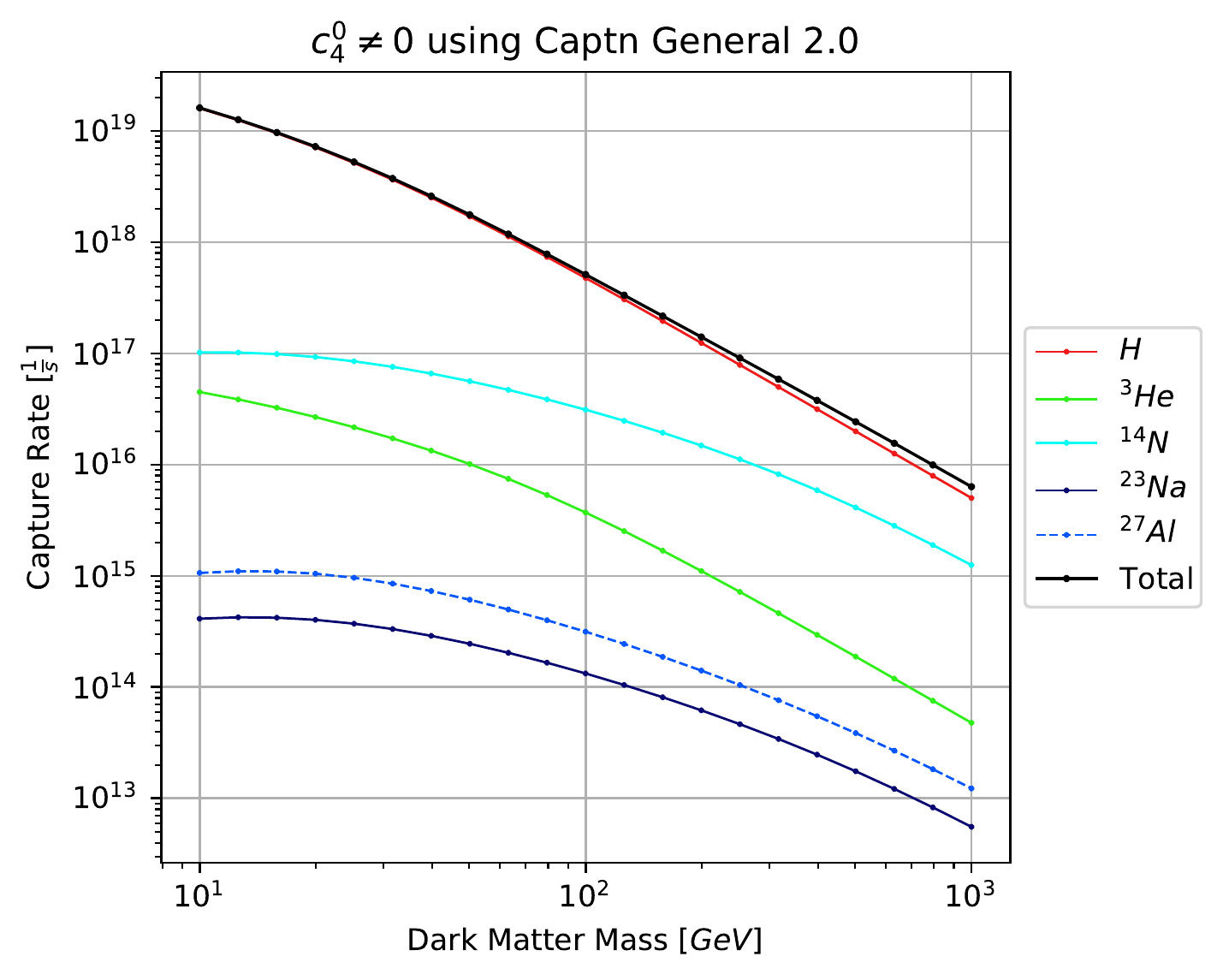}\includegraphics[width=0.5\textwidth]{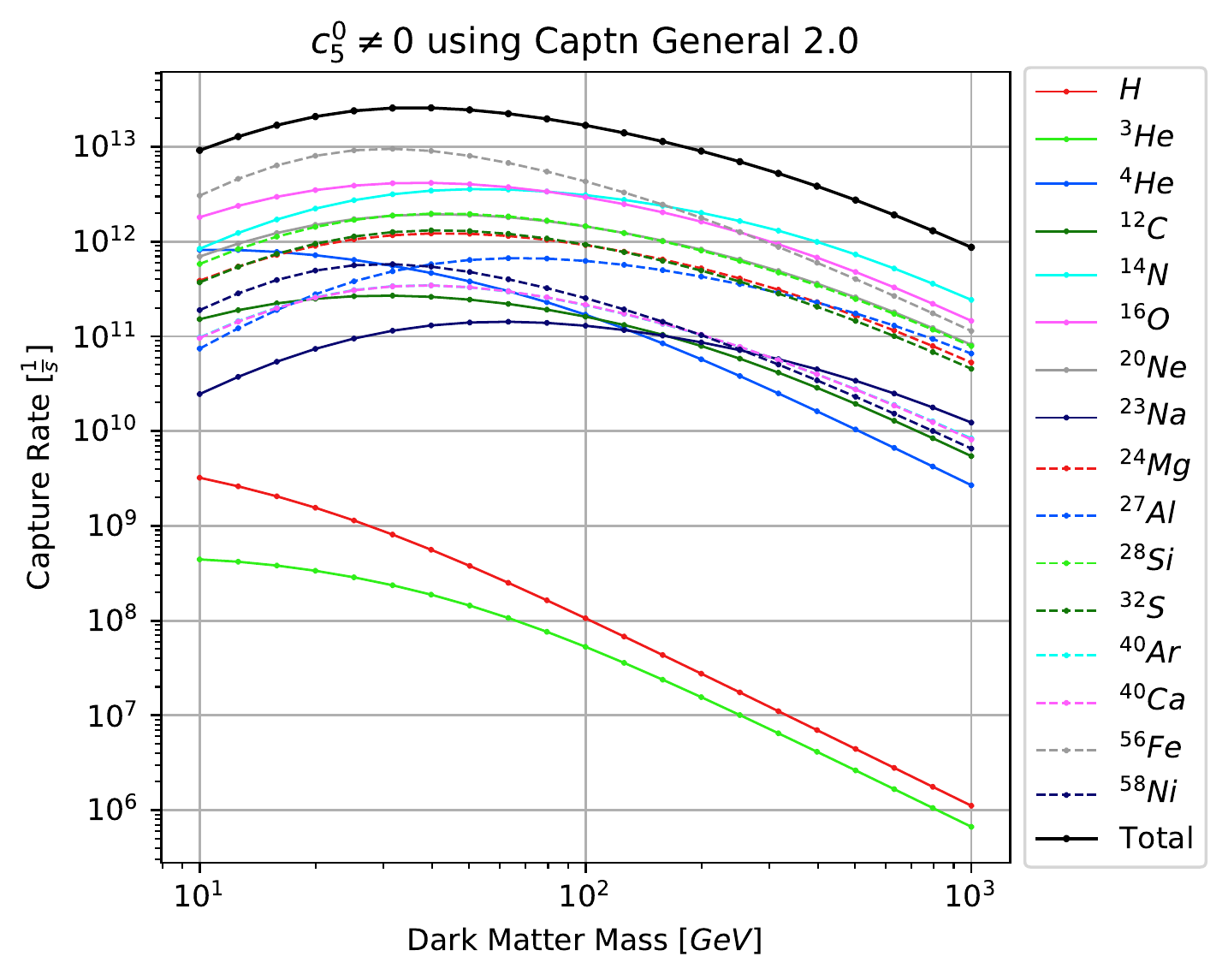}\\
    \includegraphics[width=0.5\textwidth]{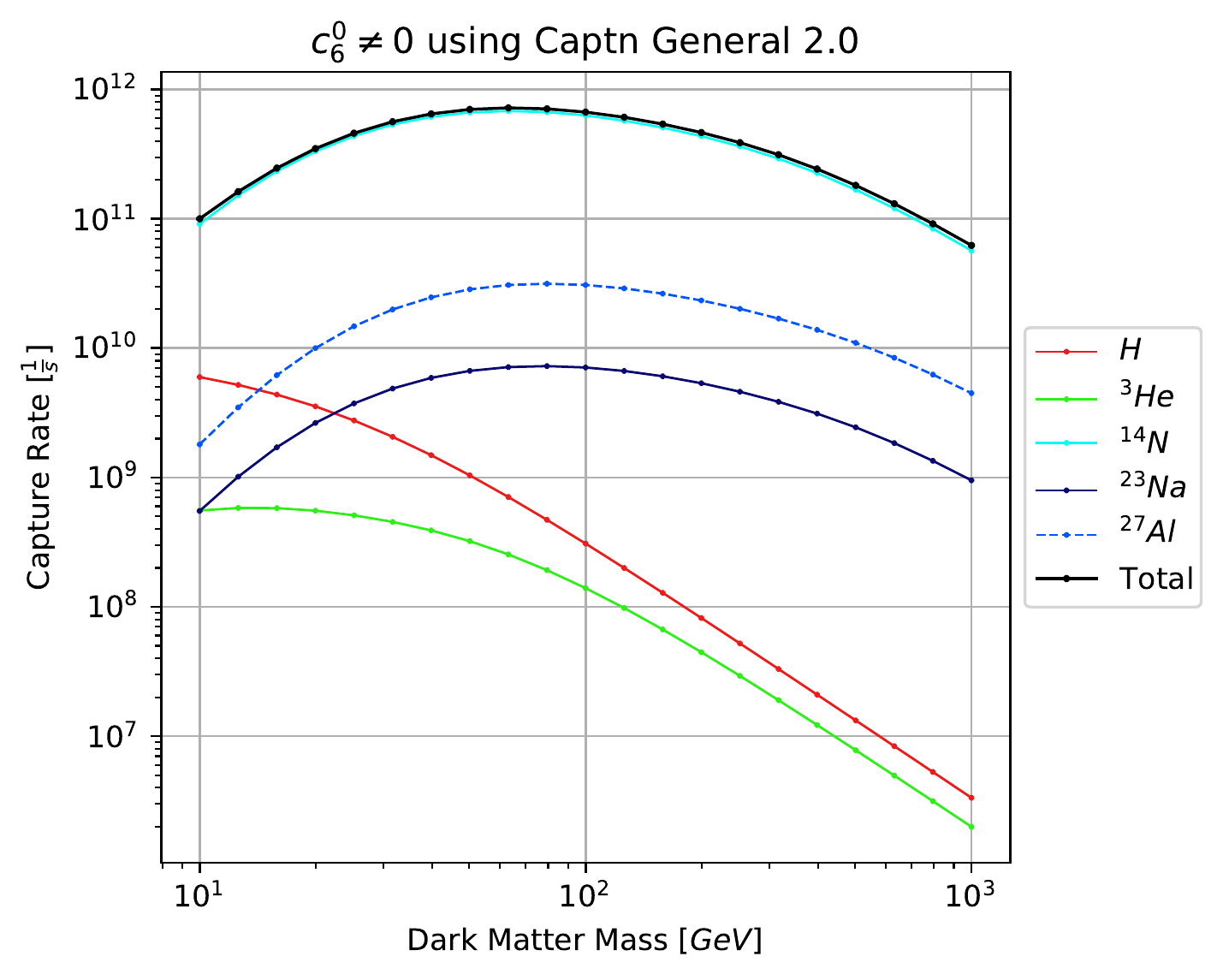}\includegraphics[width=0.5\textwidth]{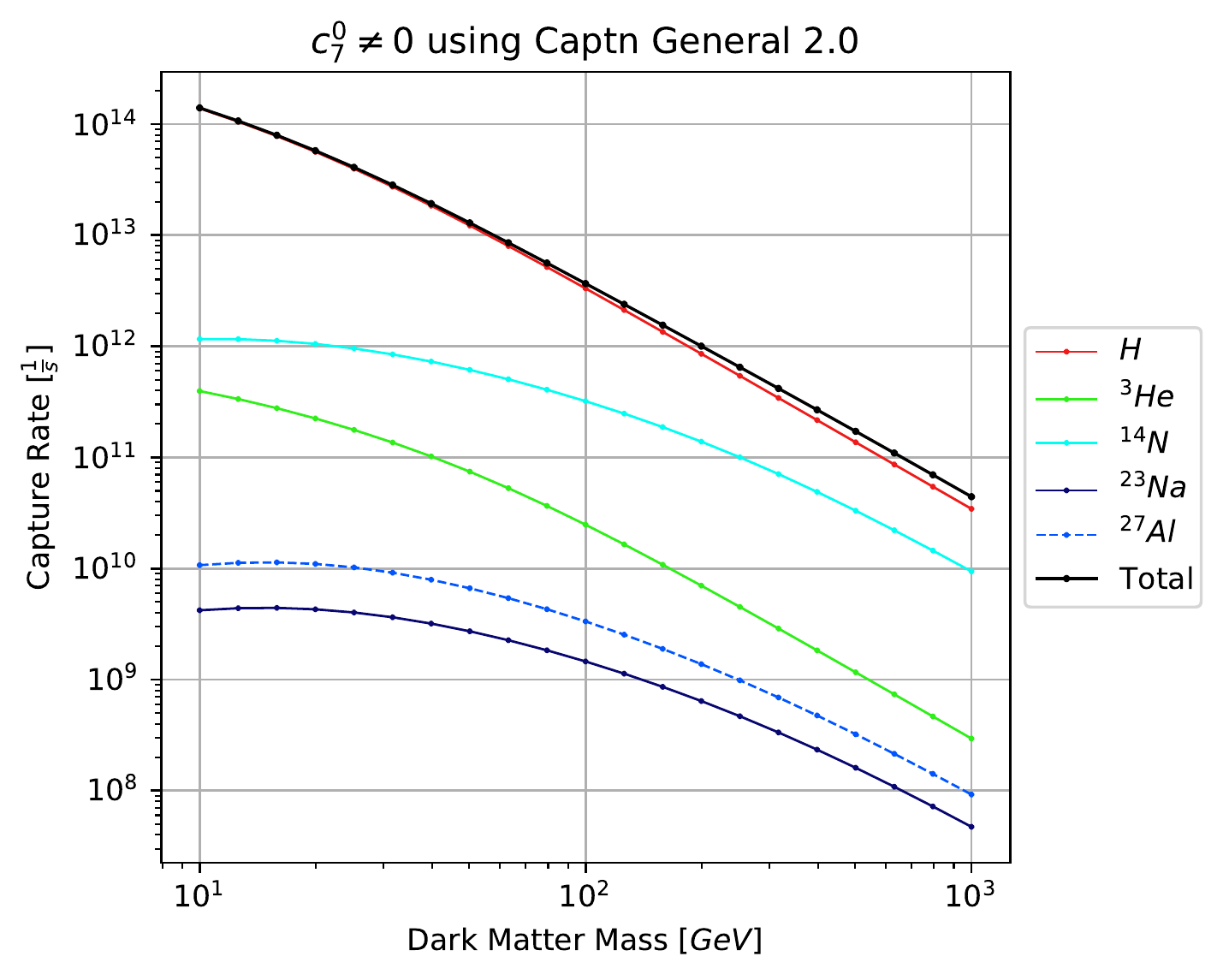}\\
    \includegraphics[width=0.5\textwidth]{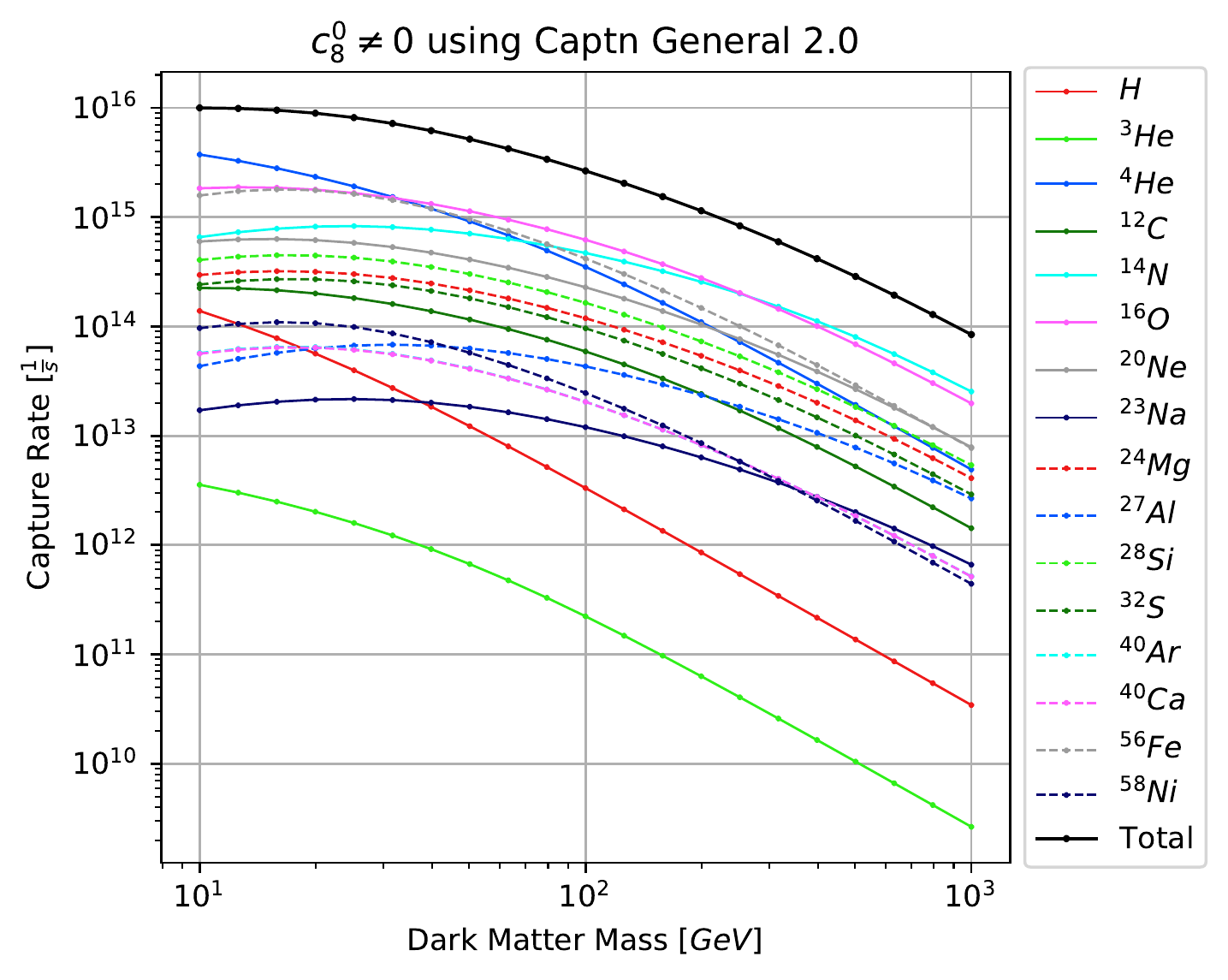}\includegraphics[width=0.5\textwidth]{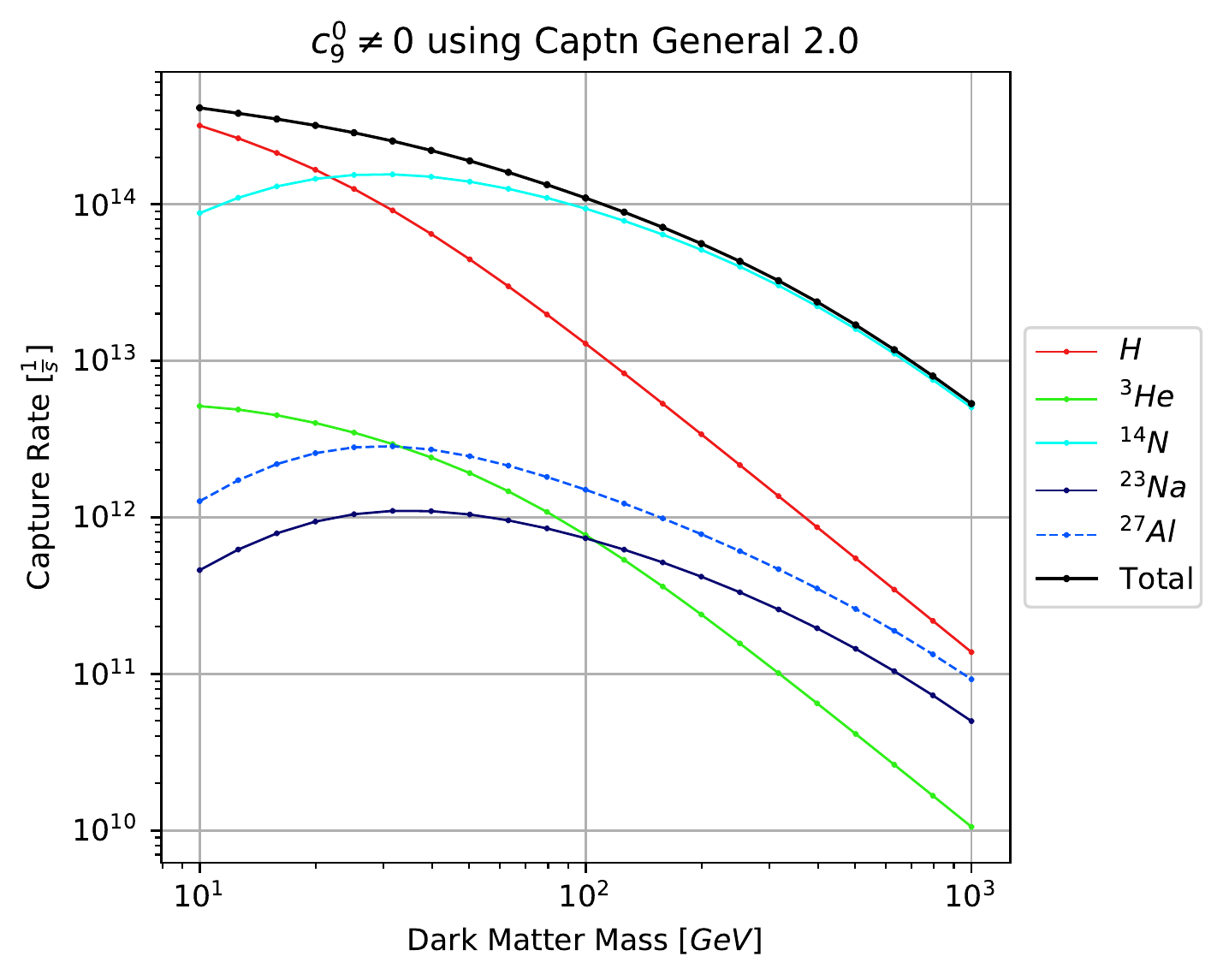}
    \caption{Capture rates of dark matter in the Sun computed with \capgen's \texttt{Capt'n Oper} nonrelativistic effective operator module for isoscalar operators $\mathcal{O}_1-\mathcal{O}_{9}$. Here, $\rho_0 = 0.4$ GeV cm$^{-3}$, $v_\odot = v_0 = 235$ km s$^{-1}$ and $v_{esc} = 550$ km s$^{-1}$. Nonzero couplings are set to $c^0_n = 1.65\times 10^{-8}$ GeV$^{-2}$.}
    \label{fig:caps1}
\end{figure}

\begin{figure}
    \centering
    \includegraphics[width=0.5\textwidth]{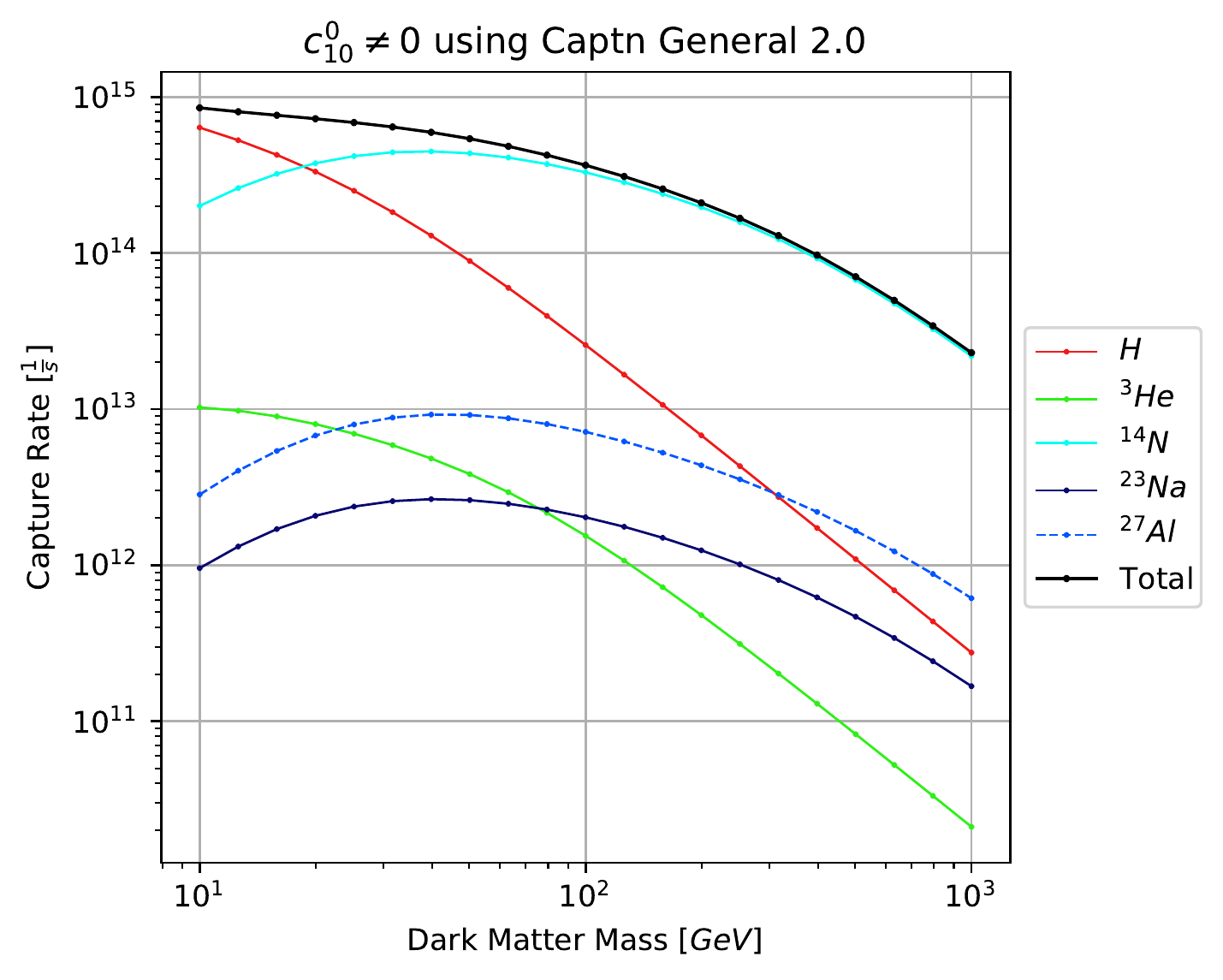}\includegraphics[width=0.5\textwidth]{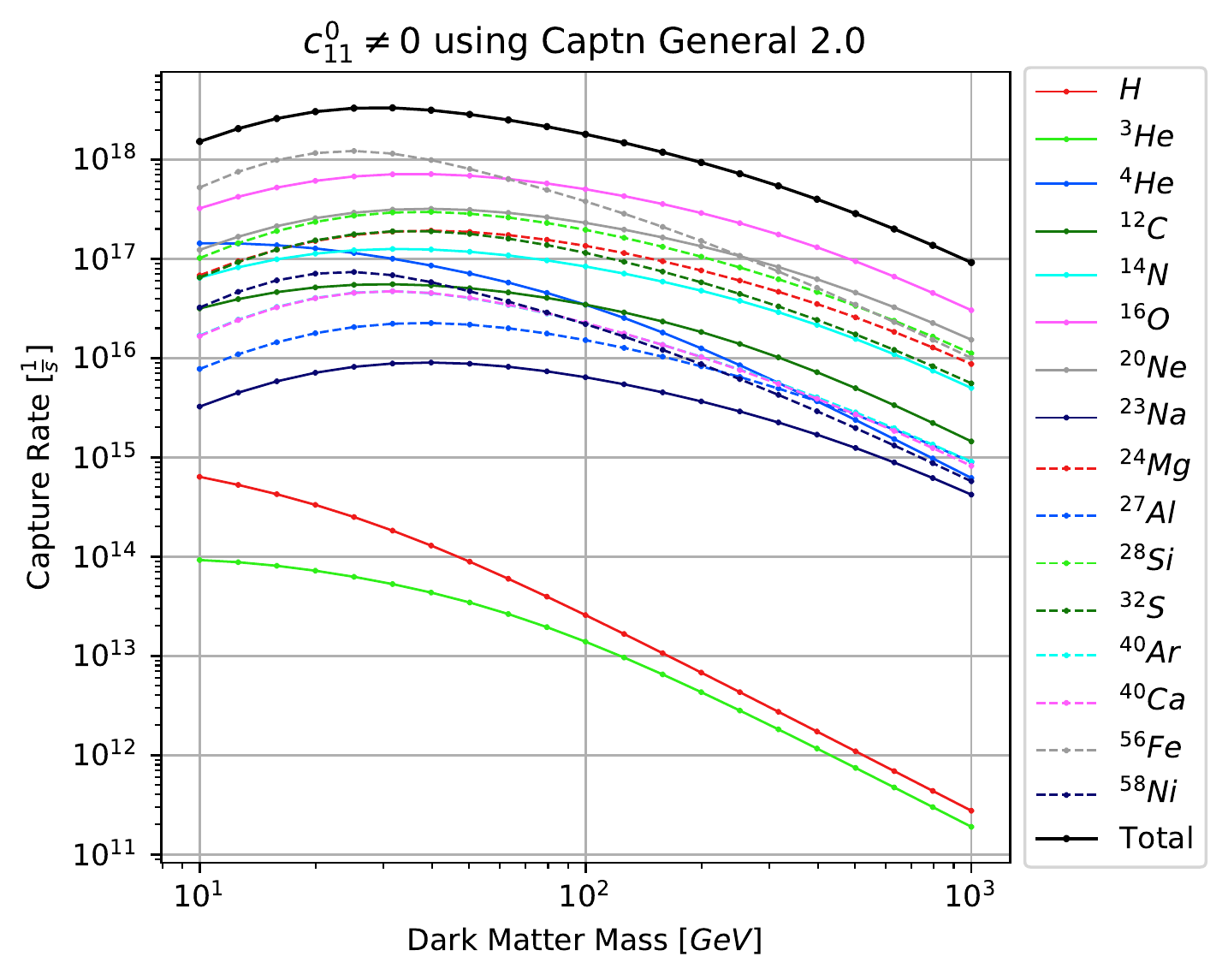}\\
    \includegraphics[width=0.5\textwidth]{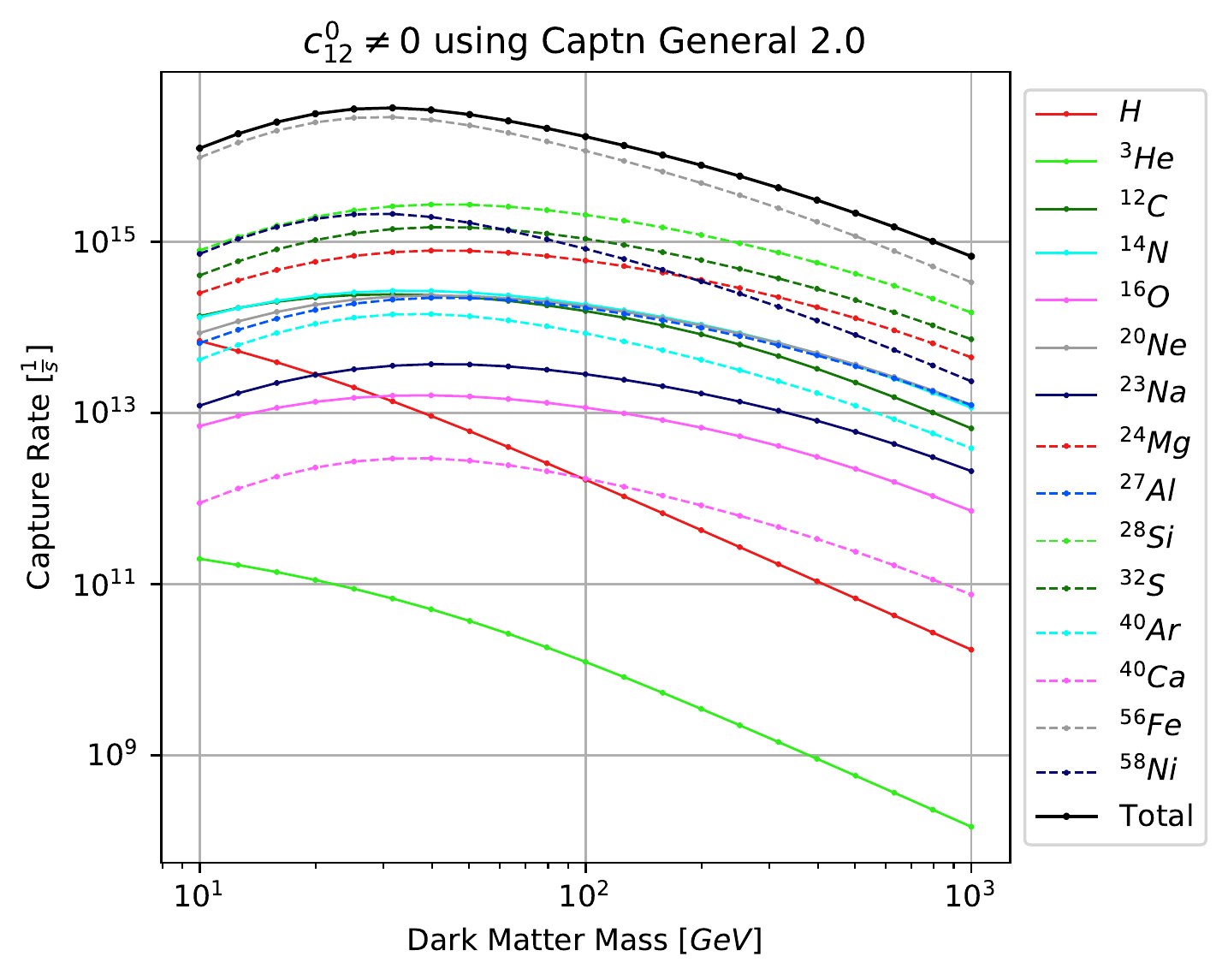}\includegraphics[width=0.5\textwidth]{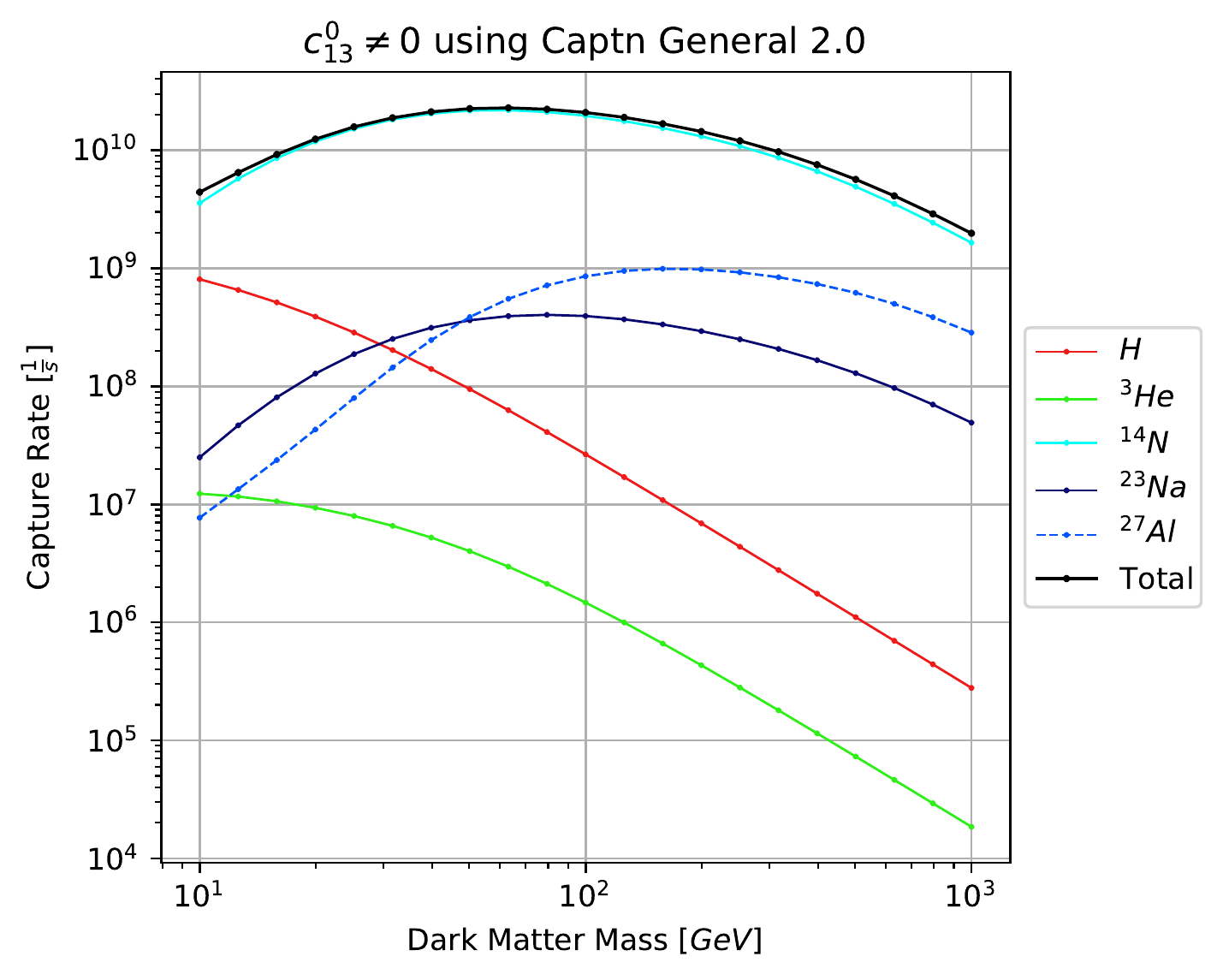}\\
    \includegraphics[width=0.5\textwidth]{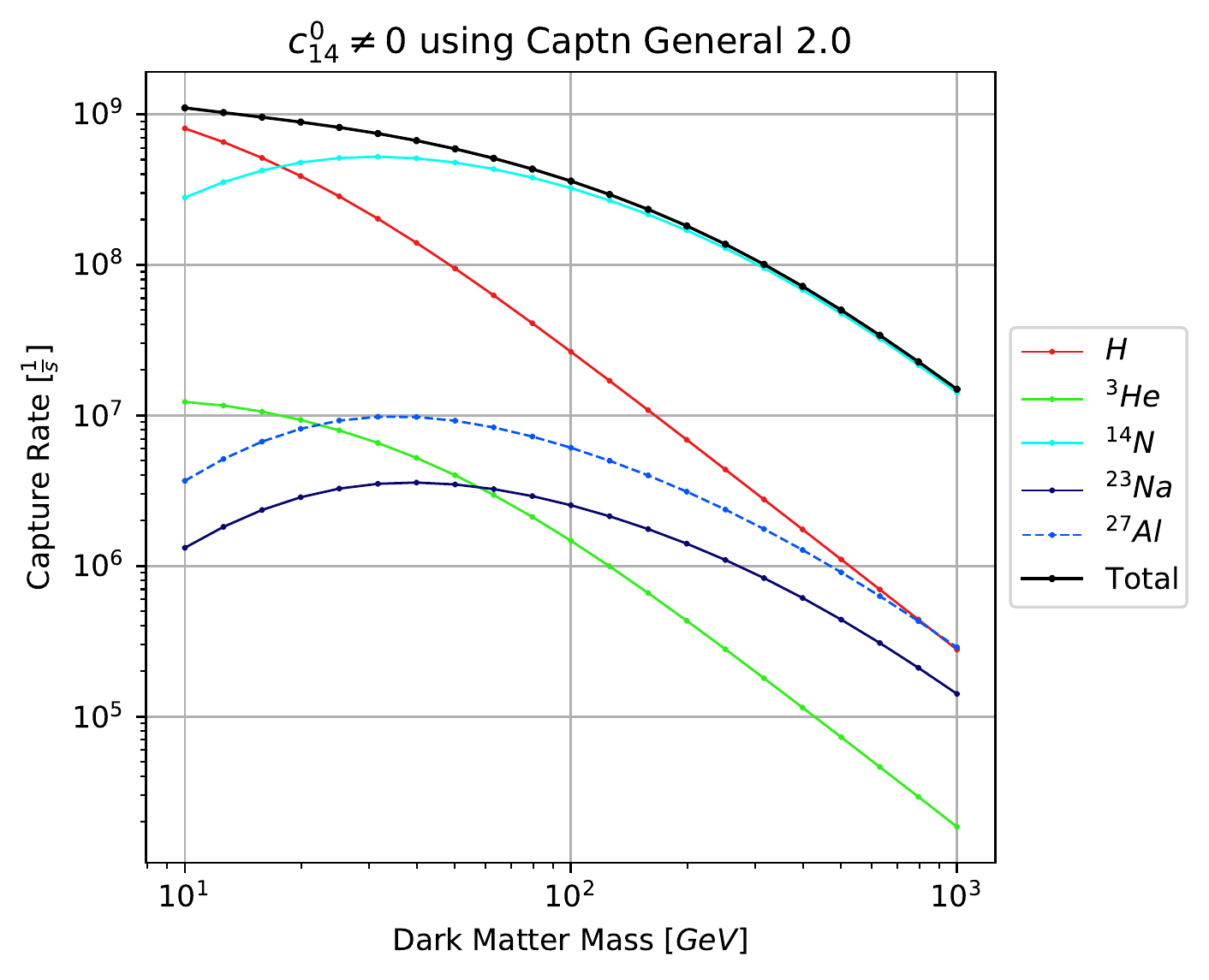}\includegraphics[width=0.5\textwidth]{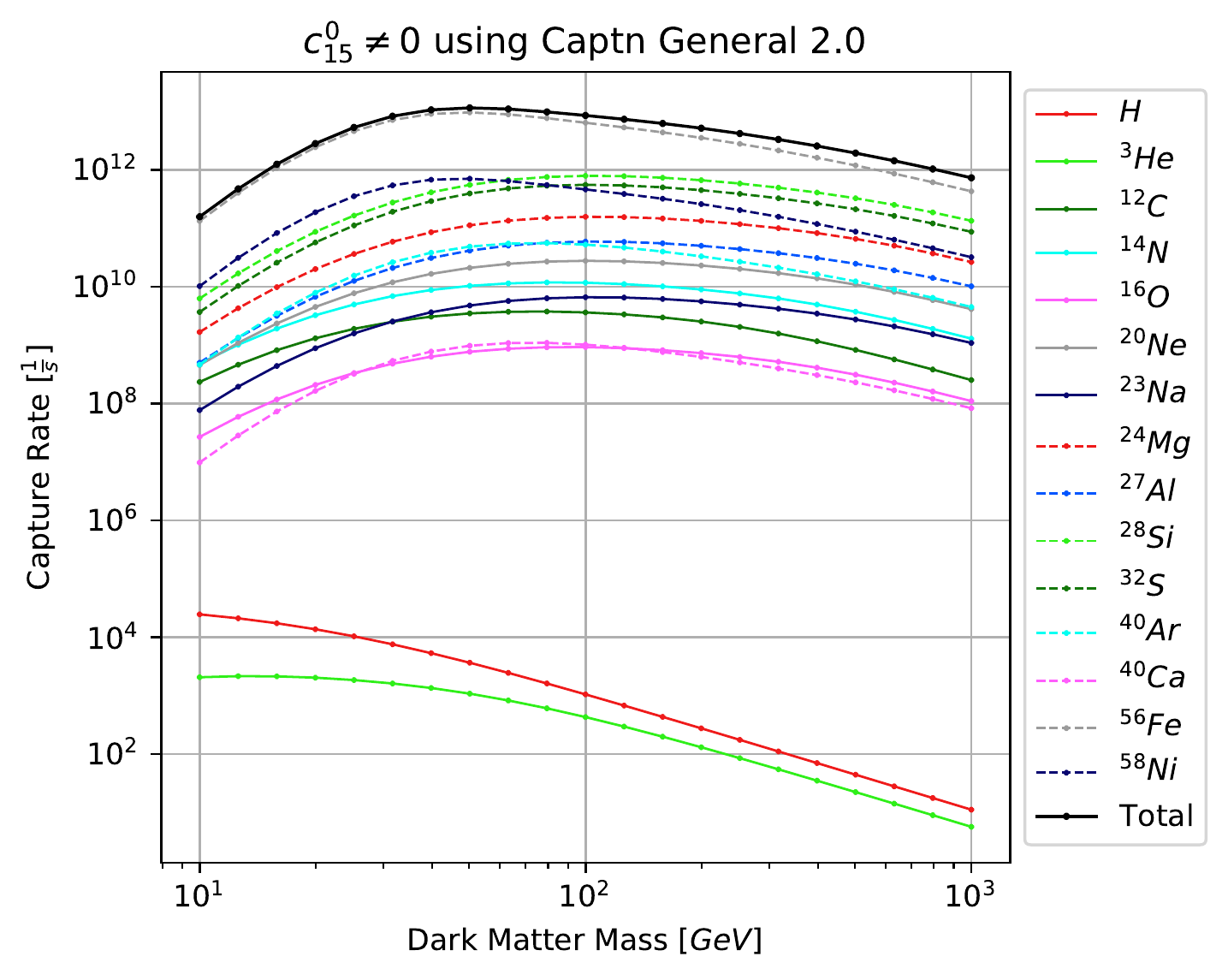}\\    
    \caption{Capture rates of dark matter in the Sun computed with \capgen's \texttt{Capt'n Oper} nonrelativistic effective operator module for isoscalar operators $\mathcal{O}_{10}-\mathcal{O}_{15}$. Here, $\rho_0 = 0.4$ GeV cm$^{-3}$, $v_\odot = v_0 = 235$ km s$^{-1}$ and $v_{esc} = 550$ km s$^{-1}$. Nonzero couplings are set to $c^0_n = 1.65\times 10^{-8}$ GeV$^{-2}$.}
    \label{fig:caps2}
\end{figure}

The nonzero $c^\tau_n$ coefficients can be populated using the \texttt{populate\_array} subroutine. Once these are specified, \texttt{captn\_oper} can compute the capture rate based on the combination of nonzero couplings. See \texttt{main.f90} for an example. Capture rates computed with this module are shown in Figs. \ref{fig:caps1} and \ref{fig:caps2}, for isoscalar coupling.

\subsection{Heat transport}
The subroutine \texttt{transgen} can be used to calculate the amount of transported heat by a population of DM in a star. By default it uses the Gould and Raffelt \cite{GouldRaffelt90a} transport formalism, extended to include transport by velocity or momentum-dependent models as in \cite{VincentScott2013}. The user/input program must provide a cross section, a number of isotopes to use, the exponents $n_q$ and $n_v$ if the interaction is momentum or velocity-dependent (as in Sec. \ref{sec:vnqn}). The simpler Spergel and Press \cite{Spergel85} method, which assumes an isothermal DM distribution, will also be available in future releases, though we caution that it overpredicts luminosity heat by (at least) an O(1) factor \cite{GouldRaffelt90a,GouldRaffelt90b}. 
In all schemes, transported energy is written out as deposited energy per unit stellar density per unit time (erg cm$^3$ g$^{-1}$ s$^{-1}$).
\subsection{Evaporation}
A full computation of the evaporation rate as a function of cross section and mass can be time-consuming, due to the importance of the optical depth to the surface at each location within the star \cite{Busoni:2017mhe}. However, the transition from negligible to full evaporation happens abruptly. The next release will therefore implement the tabulated evaporation masses as a function of stellar mass and DM-nucleon cross section computed in Ref. \cite{Garani:2021feo}.


\section{Future updates}
\label{sec:future}
As mentioned above, the \DarkMesa interface between \capgen and the stellar evolution code \mesa is in development, and will be released soon. Other plans include a Python interface, and a direct interface to directly compute annihilation and IceCube likelihoods (e.g. using the $\chi$aro$\nu$ tool \cite{Liu:2020ckq}), as well as inclusion of the multiple-scattering regime in which the optical depth becomes important for capture and evaporation \cite{Busoni:2017mhe}.

\section*{Acknowledgements}
PS acknowledges funding support from the Australian Research Council under Future Fellowship FT190100814. NAK, LFL and ACV are supported by NSERC, the Arthur B. McDonald Canadian Astroparticle Research Institute, the Canada Foundation for Innovation and the province of Ontario. Equipment was hosted and operated by the Queen's Centre for Advanced Computing.

\bibliographystyle{JHEP_pat} 
\bibliography{references.bib}

\end{document}